\documentstyle[12pt]{article}
\newcommand{\beq}{\begin{equation}}
\newcommand{\eeq}{\end{equation}}
\newcommand{\beqa}{\begin{eqnarray}}
\newcommand{\eeqa}{\end{eqnarray}}

\newcommand{\s}{{\rm s}}
\newcommand{\adj}{{\rm adj}}
\renewcommand{\d}{{\rm d}}
\newcommand{\e}{\epsilon}

\begin{document}

\begin{titlepage}

\begin{flushright}
ITP--SB--93--59
\end{flushright}

\vspace{2cm}

\begin{center}
{\bf COLOR EXCHANGE IN NEAR-FORWARD HARD ELASTIC SCATTERING }
\vspace{2cm}

Michael G. Sotiropoulos  and George Sterman \\
\vspace{1cm}
{\em  Institute for Theoretical Physics, State University of New York, \\
        Stony Brook, N.Y. 11794-3840, U.S.A.  }

\vspace{1cm}

October 12, 1993

\vspace{1.5cm}

\end{center}

\begin{abstract}

 We study the large-$t$ small angle behavior of
 quark-quark elastic scattering.
 We employ a factorization procedure previously developed
 for fixed angle scattering, which
 depends on the color structure of the
 factorized hard subprocess. We find an
 evolution in $t$ that (in leading logarithmic approximation)
 becomes diagonal in a singlet-octet
 basis in the $t$-channel as $s\rightarrow \infty$.
 Octet exhange in the hard scattering is associated with the
 familiar  `reggeized',
 $s^{\alpha_g(t)}$ behavior, which arises from $s$-dependence
 in Sudakov suppression.
 In contrast, Sudakov suppression for $t$-channel singlet exchange
 in the hard scattering is $s$-independent.
 In general, these contributions are mixed by
 soft corrections, which, however,
 cancel in many experimental amplitudes and  cross sections.
\end{abstract}

\begin{flushleft}
---------------------------------------------------- \\
{\footnotesize
email: michael@max.physics.sunysb.edu \\
email: sterman@max.physics.sunysb.edu }
\end{flushleft}

\end{titlepage}
\baselineskip=18 pt

\section{Introduction}

 Elastic scattering has been extensively studied in
 perturbation theory at fixed,
  wide angles \cite{qkctg,Land,pQCDelastic}
  and in the forward direction \cite{Reg1,Reg2}.
 Since quarks are confined, true on-shell quark-quark elastic scattering
 amplitudes are exactly zero.  Yet, in proton-proton elastic scattering
 and in jet production, we do see strong evidence for
 subprocesses involving parton-parton
 scattering.  Such a connection is possible because we
 factorize long-distance from
 short-distance behavior in the amplitude of the former,
 and the cross section of the latter.

  In this paper, we shall employ the formalism of
 wide-angle hard scattering, and push it toward the
 forward direction, $s\gg -t$, but keeping $-t\gg \Lambda^2$,
 with $\Lambda$
 the scale of the strong interactions.  In this way, we shall try
 to separate truly long-distance effects from logarithms of
 $s/t$.  Our main interest will be in the behavior of the
 short-distance part of our amplitude.

 Our techniques are closely related to those discussed by Lipatov
 in the introductory sections of
 his study of quasi-elastic unitarity \cite{Lipeu}, but our aim here is
 complementary.  Rather than study
 the role of soft gluons, we
 will focus on
  amplitudes and cross sections in which the effects
  of soft gluons cancel.
 In this respect, our results should be
 applicable to $t$-dependent corrections to jet cross
 sections of the sort investigated by Mueller and Tang \cite{Mu}
 and del Duca, Peskin and Tang \cite{DDPT}.  Another application
 is to the elastic scattering amplitude of protons at large
 $t$ and very large $s$ \cite{Land}.

 Unlike factorization in deeply inelastic scattering
 or inclusive jet production, our factorization procedure
 is carried out for fixed, rather than averaged, colors of
 the active quarks of the process.
 We will thus speak of
 the color exchange within the hard scattering, and, as we shall see
 below, we will find special, and quite different roles for
 hard `singlet' and `octet' exchange.  We must emphasize that singlet
 and octet exchange in the hard part are {\it not} the same
 things as singlet and octet exchange in the full amplitude.  Indeed,
 the color structure of the full amplitude is not infrared safe,
 since it is always possible to change octet to singlet,
 or vice-versa, by the exchange of a single zero-momentum gluon in
 the distant past or in the distant future.  Because
 our color structure will apply to short-distance
 functions, our results will be infrared safe.

 We shall rederive a difference between the asymptotic
 behavior of
 hard color singlet and color octet exchange \cite{Reg2,LM}
 in the forward limit, with the `reggeized' behavior
 of the latter arising from Sudakov suppression of elastic scattering
 at fixed angles.
 In addition, we shall see that the fixed-angle
 analysis is consistent with the summation of leading behavior in
 $\ln(s/t)$  and the simultaneous summation of leading logarithms of
 $\ln(t/\Lambda^2)$.
 Thus, we shall derive for singlet exchange in quark-quark scattering
 a result of the form
 $\exp[\ln (t)\; \ln \, \ln(t)]H_\s(s/t)$, with
 the exponential coming from Sudakov behavior in the momentum transfer,
 independent of $s$ but dependent on the infrared cutoff, and
 with $H_\s$ a hard-scattering
 function that is free of all scales below $t$.
 Because the infrared behavior has been factorized into overall
 Sudakov factors, it is possible systematically to incorporate
 our results on quark-quark scattering into hadron-hadron
 elastic scattering, as well as into inclusive cross sections involving
 singlet exchange \cite{Mu,DDPT,rapgap}.

 Wide angle and forward scattering, although related,
 differ in several important respects.
 Fixed-angle scattering is a two-scale
 problem, with the angle simply a parameter
 of order unity.  The scattering
 occurs through one \cite{qkctg} or a few \cite{Land}
 interactions that are well-localized in space-time, from which
 long-distance, nonperturbative behavior may be factorized
 into universal functions that describe the structure
 of the external hadrons in isolation \cite{pQCDelastic,BS}.  In contrast,
 near-forward scattering involves at least three scales: $s$, $t$, and
 hadronic masses.  Similarly, leading high-energy behavior is associated
 with a `ladder' structure in perturbation theory, in which there
 is a kinematic ordering, but no clearly identifiable hard scattering.
 And, in the limit of small $t$, there seems no obvious way of
 separating long- and short-distance behavior.
  Formally, we shall assume that $t$ remains large enough that
 $\alpha_s(-t)\ln (s/t)$ remains small. Although we suspect that
 our arguments can be extended beyond this restriction, we shall not do
 so here.

 The methods developed in \cite{Sen2} and \cite{BS}
 allow us to calculate
 the simpler process of
 hadron-hadron fixed-angle scattering, including leading
 and, in principle, non-leading, powers
 of $s$.  For forward scattering, most
 results are at the level of leading logarithm in $s$, although
 certain results are available for nonleading
 logarithms \cite{Lipeu,LM,Sen1}.
 In this paper, we shall study quark-quark elastic scattering
 in the formalism of ref.\ \cite{BS}, by taking the
 $\theta \rightarrow 0 $ limit, with $\theta$ the center-of-mass (CM)
 scattering angle.  We use dimensional regularization
 to handle the infrared and collinear divergences that are
 factorized, and which cancel, or are absorbed into wave functions
 when the external particles are hadrons \cite{pQCDelastic,BS}.

 In Sect.\ 2, we begin our treatment by adapting the approach of
 ref.\ \cite{BS} to dimensionally regularized quark-quark
 elastic scattering, emphasizing the role
 of angle-dependent anomalous dimensions in the
 space of color flow.  In Sect.\ 3, we examine the specific
 anomalous dimension matrix for quark-quark scattering
 calculated in ref.\ \cite{BS} in the forward-scattering limit.
 In this limit, we immediately
 rederive the leading trajectory
 for octet exchange, and show that at leading logarithm in $s$,
 hard singlet exchange remains uncorrected.
 We exhibit, however, for the singlet,
 factorized $t$ dependence, which becomes leading for large $\theta$,
 but which remains as a $t$-dependent Sudakov suppression for
 $t$ fixed, $s\rightarrow \infty$, independent of $s$.
 Sects.\ 4 and 5 discuss lowest-order singlet exchange,
 the first dealing with the one-loop diagrams
 in dimensional regularization, and the second discussing
 issues associated with the choices of finite parts in the
 factorization of short- from long-distance contributions to
 singlet exchange, treating $t$ as an ultraviolet scale, even
 though $|t|\ll s$.  We suggest a somewhat unconventional
 way of treating
 singlet exchange, in  which the lowest-order hard scattering may
 have a nonzero {\it real} contribution (we note that the remainder
 of our discussion is independent of this suggestion).
 Finally, we present our conclusions and hopes for
 future progress.

\section{Quark-quark elastic scattering at high energy}

 Consider the amplitude $A(s,t)$ for quark-quark
 elastic scattering at high energy
 and fixed CM scattering angle $\theta(s/t)$. The quarks are taken massless.
 In refs.\ \cite{Sen2,BS}, it was shown that this amplitude may be factored
 into functions that summarize the effects of: (i)
 soft gluons exchanged between quarks, (ii) lines collinear to each
 external quark, and
 (iii) infrared safe, short-distance contributions.  Let us first
 discuss some relevant details of this factorization applied
 to quark-quark scattering.

\subsection{The factorized q-q amplitude}

 The general form of a quark-quark scattering amplitude is
\beq
A_{\{ a_{i} \}}(s,t,\e;\lambda_i) = \prod_{i=1}^{4}
u_{\alpha_{i}}(P_{i},\lambda_{i})
z_{i}^{1/2}\left ({P_{i}\cdot n \over \mu\sqrt{n^2}},\e \right )
G_{ \{ \alpha_{i},a_{i} \} }
\left ({P_{i}\cdot n \over \mu\sqrt{n^2}},s,t,\e \right )\, ,
\label{qqamp}
\eeq
 where $\alpha_{i}, a_{i}$ are Dirac and ${\rm SU}(N_{c})$ indices,
 $\lambda_{i}, u_{\alpha_{i}}$ are quark helicities and spinors,
 $z_{i}$ are the residues of the full quark propagators evaluated at
 $P_{i}^{2}=0$ and $G$ is the 1PI, 4-point, on-shell vertex function.
 As indicated, $A$ is regularized in $D=4-2\epsilon$
 dimensions.  In an axial gauge the residues
 and vertex function depend on the gauge fixing vector $n^\mu$ in
 the manner shown.

 The general leading momentum region producing IR singularities in $A$
 (leading pinch singular surface )
 is shown in fig.\ 1 in the axial gauge. Here $J_{i}$ are subgraphs of on-shell
 lines nearly collinear to the external momenta $P_{i}$, $S$ is a subgraph
 attached to the $J_{i}$'s via soft gluons $q^{\mu}$ ,
 ($|q^{\mu}| \ll Q$), and $H$ is the hard subgraph whose internal lines
 are off-shell by ${\cal O}(Q^{2})$, $Q$ being the hard scale of the
 q-q scattering.
 To be specific, we shall take
\beq
Q^2=-t\, .
\label{Qdef}
\eeq
 The applicability of pQCD requires $Q^2 \gg \Lambda^2$.
 Through the use of the soft approximation and Ward identities \cite{CSS}
 the contribution of the leading region can be factorized
 in the following manner \cite{Sen2,BS}.

 First, soft gluon connections, $S$, between jets are factored
 into an eikonal function
 $U_{ \{ a_{i},b_{i} \} }(\alpha_s(\mu),\mu,\theta,\epsilon)$ that carries
 no Dirac structure and can be expanded as a series in the coupling constant
\beq
U_{ \{ a_{i},b_{i} \} }=\prod_{i=1}^{4} \delta_{a_{i}b_{i}} +
{\cal O}(\alpha_s)\, .
\label{Uphase}
\eeq

 Next, the remaining connections of $S$ that are
 one-particle reducible are included in the jet functions
 $J_{i}((P_{i}\cdot n/\mu\sqrt{n^2}),\epsilon)$, where
 $\mu$ is the renormalization scale and $v_{i}$ are light-like vectors along
 the direction of motion of the external quarks,
\beq
P_{i}^\mu=
v_{i}^\mu\sqrt{s/2}\, .
\label{videf}
\eeq
  In an arbitrary gauge, $n^\mu$ may be thought of as a space-like vector
 introduced to define the jet function as a specific
 matrix element in the manner described in ref.\ \cite{CSrv}.
 In the $n\cdot A=0$ gauge, however, the jet functions $J$
 are identical to the residues $z_i^{1/2}$ in eq.\ (\ref{qqamp}).
 To be specific, we shall work in axial gauge, but we
 emphasize that our results are  gauge-independent.

 With collinear and infrared contributions organized into
 the functions $z_i$ and $U$, the remaining factor, denoted $H$
 below, is short-distance dominated. This separation is
 illustrated in  fig.\ 2.
 In summary, the factorized form of the amplitude in axial gauge
 may be written as in eq.\ (\ref{qqamp}), with
 the vertex function $G$ given by
\beq
G_{ \{ \alpha_{i},a_{i} \}}(Q/\mu,\alpha_s(\mu),\theta,\epsilon)=
U_{ \{ a_{i},b_{i} \}}(\alpha_s(\mu),\theta,\epsilon)
H_{ \{ \alpha_{i},b_{i} \}}(Q/\mu,\alpha_s(\mu),\theta)\, .
\label{factorizationG}
\eeq
 Here, we have replaced the $P_i$- and $n$-dependence in $G$ by
 $Q$ and $\theta$.  Again, in a more general gauge, the $z_i$ of
 eq.\ (\ref{qqamp}) would be replaced by a jet function $J_i$
 which would equal the $z_i^{1/2}$ of the $n\cdot A=0$ gauge.
 Note that $U$, although angular-dependent (through $v_i\cdot n$),
 depends on $\mu$ only through $\alpha_s$, and is otherwise
 independent of the mass scales of the problem.

 As discussed in ref.\ \cite{BS}, the functions $U$ and $H$
 must be defined through a renormalization procedure \cite{CS}.
 To this end, it is now convenient to reexpress the color sums
 in eq.\ (\ref{factorizationG}) in terms of a specific basis.
 As we see, this will also enable us to treat the $Q$-dependence
 of these functions by renormalization group methods.

\subsection{Color flow and renormalization}

 Color flow in quark-quark scattering may always be
 decomposed according to a color basis $c_{_{I}}$,
 defined by
\beq
(c_1)_{\{ a_{i} \}}=\delta_{a_{1}a_{4}}\delta_{a_{2}a_{3}}
\mbox{\, ,\hspace{1cm}}
(c_2)_{\{ a_{i} \}}=\delta_{a_{1}a_{3}}\delta_{a_{2}a_{4}}\, .
\label{cdef}
\eeq
 In terms of this basis, we may write
\beq
A_{ \{ a_{i} \}}=A_{_{I}}\left ( c_{_{I}}\right )_{ \{ a_{i} \}} \, ,
\hspace{.25cm}
G_{ \{ a_{i} \}}=G_{_{I}}\left ( c_{_{I}}\right )_{ \{ a_{i} \}} \, ,
\hspace{.25cm}
H_{ \{ b_{i} \}}=H_{_{I}}\left (c_{_{I}}\right )_{ \{ b_{i} \}}  \, ,
\label{AGHcolor}
\eeq
 and we define
\beq
U_{ \{ a_{i},b_{i} \}} \left (c_{_{J}}\right )_{\{ b_{i} \}}
\equiv U_{_{IJ}} \left (c_{_{I}}\right )_{\{ a_{i} \}}\,.
\label{colormix}
\eeq
Then the vertex function can be writen as
\beq
G_{_{I}}=U_{_{IJ}}H_{_{J}}\, .
\label{vertexcolor}
\eeq
 with perturbative expansion for $U_{_{IJ}}$ as
\beq
U_{_{IJ}}=\delta_{_{IJ}}+{\cal O}(\alpha_s)\, .
\label{upert}
\eeq
 From eqs.\ (\ref{colormix}) and (\ref{vertexcolor}) it is clear
 how soft gluon insertions lead to color mixing.
 The matrix $U_{_{IJ}}$ is generated by dressing
 the `hard' color tensor $c_{_{J}}$ with soft gluons and
 projecting the result along the direction of $c_{_{I}}$.

 We now discuss the renormalization of the eikonal and
 hard scattering functions in the above color basis.  We denote by
 $U^{(0)}_{_{IJ}}$ and $H^{(0)}_{_{I}}$the UV-divergent
 unrenormalized soft and hard functions.  Their
 multiplicative renormalization \cite{BS}
 involves the same color mixing as in the amplitude,
\begin{eqnarray}
U^{(0)}_{_{IJ}}(\alpha_s(\mu),\theta,\epsilon)&=&
U_{_{IK}}(\alpha_s(\mu),\theta,\e)
\left (Z_U \right )_{_{KJ}}(a_s(\mu),\theta,\epsilon) \, , \nonumber \\
H^{(0)}_{_{I}}(Q/\mu,\alpha_s(\mu),\theta,\epsilon)&=&
\left( Z_\psi ^{-1} \right)^4
\left ( Z^{-1}_U\right )_{_{IJ}}(a_s(\mu),\theta,\epsilon)
H_{_{J}}(Q/\mu,\alpha_s(\mu),\theta)\, .
\label{UHrenorm}
\end{eqnarray}
 Hence, the renormalized $H$ and
 $U$ satisfy the complementary matrix renormalization group equations
\beq
{\cal D}U_{_{IJ}}=-U_{_{IK}}\left ( \Gamma_U \right )_{_{KJ}} \, ,
\hspace{.4cm}
{\cal D}H_{_{I}}= \left[ \left ( \Gamma_U \right )_{_{IK}} +
                        4\gamma_\psi \delta_{_{IK}} \right] H_{_{K}} \, ,
\label{UHrge}
\eeq
 with
\beqa
{\cal D} &=& \mu{\partial \over \partial \mu} + \beta(g) {\partial
\over \partial g}\, , \nonumber
\label{Ddef}
\eeqa
 and
\beq
\gamma_\psi(\alpha_s(\mu))=-\frac{\alpha_s}{4\pi} N_c C_F
 +{\cal O}(\alpha_s^2)
\label{gammapsi}
\eeq
 the anomalous dimension of the quark wave-function renormalization
 in the axial gauge \cite{Soper}.
 For minimal subtraction and $D=4-2\epsilon$ the anomalous dimension matrix
${\bf\Gamma}_{U}$ is given simply by
\beq
{\bf \Gamma}_{U}(\alpha_s(\mu),\theta)=
-g\frac{\partial}{\partial g}{\rm Res}
{\bf Z}_{U}(\alpha_s(\mu),\theta,\epsilon)\, .
\label{residue}
\eeq
 We note that because, as mentioned above, $U$ depends on no momentum
 scales, the unrenormalized $U^{(0)}$ actually vanishes in
 dimensional regularization beyond zeroth order, eq.\ (\ref{upert}).
 As a result, the renormalized $U$ is given by $Z_U^{-1}$,
\beq
U_{_{IJ}}= \left ( Z^{-1}_U \right )_{_{IJ}} \, .
\label{uequalz}
\eeq
 We are now ready to discuss the energy dependence of the
 amplitude in terms of its factorized form.

\subsection{Energy dependence}

 At lowest order in $\alpha_s$
 the diagonalization of of the matrix ${\bf\Gamma}_{U}$ may be performed
 with a matrix ${\bf R}$ that is independent of $\mu$ and $\alpha_s(\mu)$.
\beq
({\bf R}^{-1}{\bf \Gamma}_{U}{\bf R})_{_{IJ}}=
\lambda_{_{I}}(\alpha_s(\mu),\theta)\delta_{_{IJ}},
\hspace{.5cm}
\hat{U}_{_{IJ}}=R^{-1}_{_{IK}}U_{_{KL}}R_{_{LJ}},
\hspace{.5cm}
\hat{H}_{_{J}}=R^{-1}_{_{JK}}H_{_{K}} \, .
\label{Gamdiag}
\eeq
 In the new basis we find decoupled renormalization group
 equations (\ref{UHrge}) for the components of both $H$ and $U$
 in the space of color flow.
 We emphasize that the eigenvalues $\lambda_{_{I}}$ are in general
 $\theta$-dependent.

 To use the information contained in eqs.\ (\ref{UHrge}), we
 recall that the quark-quark scattering amplitude should
 be thought of as embedded in a physical process,
 involving hadron-hadron elastic
 or inelastic scattering.  In any such process,
 the infrared divergences of the eikonal will cancel.  For
 elastic scattering this will occur through the color
 singlet nature of the external hadrons; in inclusive
 cross sections, it will occur through cancellation between
 real and virtual processes in perturbation theory.  In
 either case, the divergences of the soft gluon function $U$
 will be replaced by a
 dynamically-generated (Sudakov)
 transverse momentum scale
  \cite{BS} or by an energy resolution
 characteristic of the inclusive cross section \cite{CSS}.
 With this in mind, we shall use eq.\ (\ref{UHrge}) to
 evolve the scale of the coupling in $U$ from $\mu$ down to an
 `infrared' scale that we shall denote by $1/b \gg \Lambda$, with $b$
 a length,
 referring to the notation of ref.\ \cite{BS}.
 Similarly,
 for $H$, we evolve from the general scale $\mu$ up to the
 scale $Q$, eq.\ (\ref{Qdef}).
 Following this procedure, we may write the
 vertex function $G$ in eq.\ (\ref{factorizationG}) in
 terms of the leading order $\lambda_{_{I}}$ as
\beqa
&\ &\hat{G}_{_{I}}(Q/\mu,\alpha_s(\mu),\theta,\epsilon)\hat{c}_{_{I}} =
\exp \left[ -\frac{2 N_c C_F}{\beta_1}
     \ln \frac{ \ln (C_2 Q/ \Lambda)}{\ln (\mu / \Lambda)} \right]
\nonumber \\
&\ & \times \sum_J
\exp \left[-\gamma_{_{J}}(Q,b,\theta) \right]
\hat{U}_{_{IJ}}(\alpha_s(C_{1}/b),\theta,\e)
\hat{H}_{_{J}}(1/C_2,\alpha_s(C_{2}Q),\theta) \hat{c}_{_{I}} \, ,
\label{evolution}
\eeqa
with $\hat{c}$ the basis in which ${\bf \Gamma}_{U}$ is diagonal,
and with exponent given by
\beq
\gamma_{_{J}}(Q,b,\theta) =
\int_{C_{1}/b}^{C_{2}Q} \frac{\d \mu^{\prime}}{\mu^{\prime}}
\lambda_{_{J}}(\alpha_s(\mu^{\prime}),\theta)  \, .
\label{gammadef}
\eeq
The constants $C_{1},C_{2}$ are order unity,
chosen so as to optimize perturbative calculations \cite{CS}.

 We denote the expansion of the eigenvalues as a series
 in the coupling constant as
\beq
\lambda_{_{I}}(\alpha_s(\mu),\theta) = \sum_{n}
\left( \frac{\alpha_s}{\pi} \right)^{n} \lambda_{_{I}}^{(n)}(\theta) \, .
\eeq
Then from the 1-loop eigenvalues, with
\beq
\alpha_s(\mu)=\frac{4 \pi}{\beta_{1} \ln(\mu^{2}/ \Lambda^{2})}
\mbox{, \hspace{1cm}}
\beta_{1}= \frac{11}{3}N_{c}-\frac{2}{3}n_{f}\, ,
\eeq
we obtain from eq.\ (\ref{evolution}) the behavior
\beqa
\hat{G}_{_{I}}(Q/\mu,\alpha_s(\mu),\theta,\epsilon)\hat{c}_{_{I}} &=&
\exp \left[ -\frac{2 N_c C_F}{\beta_1}
     \ln \frac{ \ln (C_2 Q/ \Lambda)}{\ln (\mu/ \Lambda)} \right]
\nonumber \\
&\ & \times
\sum_J
\exp \left[-
\frac{2 \lambda_{_{J}}^{(1)}(\theta)} {\beta_{1}}
\ln \left( \frac{\ln(C_{2}Q/ \Lambda)}{\ln(C_{1}/ b \Lambda)} \right)
\right]
\nonumber \\
&\ & \times
\hat{U}_{_{IJ}}(\alpha_s(C_{1}/b),\theta,\e)
\hat{H}_{_{J}}(1/C_2,\alpha_s(C_{2}Q),\theta) \hat{c}_{_{I}} \, .
\label{evolsoln}
\eeqa
 Here the $Q$-dependence is determined by the color flow in the
 {\it hard} scattering.  $U_{_{IJ}}$, which summarizes the
 infrared behavior, may still mix colors and depend on the
 scattering angle.  We shall assume that our q-q scattering amplitude
 will appear in an IR safe cross section in which this extra
 dependence cancels.

 In order to complete the picture we must include
 the energy dependence of
 the four jet functions, $z_i$ in eq.\ (\ref{qqamp}).
 As in the case of the eikonal function $U$, we appeal to
 the eventual incorporation of the quark-quark amplitude into
 a physical process, and evolve the $z_i$ from scale $\mu$
 to an infrared scale denoted, as above, by $1/b$.
 This may be done in precisely the manner outlined in
 ref.\ \cite{BS}, with the result,
\beqa
z_i^{1/2}(\sqrt{s \nu_i/\mu^2},b\mu,\alpha_s(\mu), \epsilon) &=&
\exp \left[ -\frac{2 C_{F}}{\beta_{1}}
             \ln(\sqrt{2 s\nu_i/\Lambda^2}) \ln \left(
           \frac{ \ln(\sqrt{2 s\nu_i/ \Lambda^2})}
             {\ln (C_1/b \Lambda)} \right) + {\rm NL }
    \right] \nonumber \\
&\ &\times z_i^{1/2}(1,1,\alpha_s(C_1/b),\epsilon) \, ,
\label{zevol}
\eeqa
where
\beq
\nu_{i}=\frac{(v_{i}\cdot n)^{2}}{|n^{2}|} \, ,
\label{nudef}
\eeq
 with $n$ the axial gauge vector chosen spacelike, $n^{2}=-1$.
 `NL' stands for non-leading logarithmic corrections coming from
 the RG evolution of $z_i$ ~\cite{BS}  that we shall not need explicitly.
 The $\mu$-dependence of $z_i^{1/2}$ is now in the
 contribution to the exponent from the quark anomalous dimension,
 eq.\ (\ref{gammapsi}),
 and is also contained in `NL'. Of course,
 $\prod_i z_i^{1/2} G_{_I}$ is RG invariant.

 Combining eqs.~(\ref{qqamp}), (\ref{evolsoln})
 and (\ref{zevol}), we may summarize the energy behavior of the
 quark-quark scattering amplitude as
\beq
\hat{A_{_{I}}} = \hat{S}_{_{IJ}}(\alpha_s(1/b), \theta, \epsilon)
{\cal H}_{_{J}}(\alpha_s(Q),\theta, b;\lambda_i)
\label{abehave}
\eeq
where
\beqa
\hat{S}_{_{IJ}} &\equiv& \prod_{i=1}^4 z_i^{1/2}(1,1,\alpha_s(1/b),\epsilon)
\hat{U}_{_{IJ}}(\alpha_s(1/b),\theta,\epsilon)\, , \\
{\cal H}_{_J} &\equiv&
  \exp \left[ -\sum_{i=1}^4 \frac{2 C_{F}}{\beta_{1}}
  \ln(\sqrt{2 s\nu_i/\Lambda^2}) \ln \left(
 \frac{ \ln(\sqrt{2 s\nu_i/\Lambda^2}} {\ln (1/b \Lambda)} \right)
+{\rm NL} \right] \hat{H}_{_{J}}(1,\alpha_s(Q),\theta;\lambda_i) \, .
\nonumber
\label{calH}
\eeqa
\newpage

 We set $C_1=C_2=1$ for simplicity, and from now on we include
 the Dirac spinors in $H$.
 We distinguish between the IR regulator $\epsilon$ and the factorization
 scale $1/b$, which, as noted above, is a physical IR cutoff
 characteristic of the experiment.
 $\hat{S}_{IJ}$ contains all the IR structure of the q-q amplitude
 whereas ${\cal H}_{J}$ is the Sudakov-evolved hard part that enters
 hadronic cross sections.
 For scattering at fixed angles,
 the leading logarithms are generated by the self energies,
 through the $\nu_i$.  We shall see below, however, that  for
 the hard octet component of
 forward scattering, the leading logarithms in $s/t$ are generated
 through the anomalous dimensions of the vertex function $G=UH$.

\subsection{One-loop anomalous dimensions}

 We close this section with a summary of the one-loop anomalous
 dimension matrix ${\bf \Gamma}_U$ in the  basis defined by
 eq.\ (\ref{cdef}).

 The matrix ${\bf \Gamma}_{U}$ has been calculated \footnote{These
 results correct a misprint in ref.\ \cite{BS}, in which the
 (12) and (21) matrix elements of $\Gamma_U^{(3)}$ are too small by
 a factor of 2.} to
 ${\cal O}(\alpha_s)$ in ref.\ \cite{BS}. In terms of kinematic invariants
 it is
\beqa
\Gamma_{U 11} &=& \frac{\alpha_s}{\pi}  \left \{
C_{F} \left[ \ln  \left(  \frac{u^{2}}{4s^{2}}  \right)
             -\frac{1}{2} \ln \prod_{i=1}^{4} \nu_{i} +2  \right]
-\frac{1}{2N_{c}}\left[ \ln \left( \frac{t^{2}}{s^{2}} \right)
                        +2\pi i \right]  \right \}\, ,  \nonumber \\
\Gamma_{U 12} &=& \frac{\alpha_s}{\pi} \frac{1}{2} \left[
        \ln \left( \frac{u^{2}}{s^{2}} \right) +2 \pi i \right]\, ,
        \nonumber \\
\Gamma_{U 21} &=& \Gamma_{U 12}|_{t\leftrightarrow u}\, ,
\hspace{.3cm}
\Gamma_{U 22} = \Gamma_{U 11}|_{t\leftrightarrow u}\, .
\label{Goneloop}
\eeqa
 Here the ${\rm SU}(N_{c})$ normalization is, as usual,
\beq
{\rm Tr} \{ t_{m} t_{n} \}=\frac{1}{2} \delta_{mn},
\hspace{.5cm}
C_{F}=\frac{N_{c}^{2}-1}{2N_{c}}\, .
\label{groupnorm}
\eeq
 ${\bf \Gamma}_{U}$ is gauge dependent via the combination
 $\ln\prod_{i=1}^{4} \nu_{i}$, where $\nu_i$ has been defined in
 eq.\ (\ref{nudef}). In the following section, we shall use these results
 to study scattering in the limit of fixed $t$, $s\rightarrow \infty$.
 \newpage

\section{ Color flow in the forward direction}

 In the kinematic region of forward elastic scattering with large momentum
 transfer, $\ln(s/-t) \gg 1$,
 the eigenvalues of ${\bf \Gamma}_{U}$ are
\beqa
\lambda_{1}^{(1)} &=& \frac{1}{N_{c}}
\left[ \ln \left( \frac{s}{-t} \right) - 2 \pi i \right]
+ 2C_{F} \left[ 1-\ln2- \frac{1}{4} \ln \prod _{i=1}^{4}\nu_{i} \right]
+ {\cal O}(\ln^{-1}(s/t))\, , \nonumber \\
\lambda_{2}^{(1)} &=&  2C_{F} \left[  -\ln \left( \frac{s}{-t} \right)
 + 1-\ln2 - \frac{1}{4} \ln \prod _{i=1}^{4}\nu_{i}  \right]
+ {\cal O}(\ln^{-1}(s/t))\, .
\eeqa
 We observe that their difference is gauge independent:
\beq
\Delta \lambda^{(1)} \equiv \lambda^{(1)}-\lambda^{(2)} =
N_{c} \ln \left( \frac{s}{-t} \right) - \frac{2 \pi i}{N_{c}}
+ {\cal O}(\ln^{-1}(s/t))\, .
\eeq
 Using this result in eq.\ (\ref{evolsoln}) for
 $\hat{G}_{_I}$, the component of the vertex function $G$ along the
 diagonalized color
 channel $\hat{c}_1$ is (Sudakov) suppressed relative to the component
 along $\hat{c}_2$ by a factor
\beq
\exp\left[ - \frac{2\Delta\lambda^{(1)}}{\beta_1}
          \ln \left( \frac{\ln (Q/ \Lambda)}{\ln (1/ b\Lambda)}
             \right)
   \right]
\approx
\left( \frac{s}{-t} \right) ^{-\frac{2 N_{c}}{\beta_{1}}
\ln \left( \frac{\ln (Q/ \Lambda)}{\ln (1/ b\Lambda)} \right)}\, .
\label{suppress}
\eeq
The eigenvectors are  also gauge independent, and in the
$c_{_{I}}$ basis of eq.\ (\ref{cdef}) they are
\beq
e_{1}=\frac{1}{2} \left( \begin{array}{c} 1 \\ -1/N_{c} \end{array} \right)
+{\cal O}(\ln^{-1}(s/t))\, ,
\hspace{1cm}
e_{2}=\left( \begin{array}{c} 0 \\ 1 \end{array} \right)
+{\cal O}(\ln^{-1}(s/t))\, ,
\label{edef}
\eeq
where the  factor $1/2$ is a convenient normalization. In
the $s\rightarrow \infty$ limit, the matrix
${\bf R}=(e_{1},e_{2})$ generates the diagonaled basis $\hat{c}_{_{I}}$
\beq
\hat{c}_{1}= (t_{m})_{a_{3}a_{1}}(t_{m})_{a_{4}a_{2}} \equiv c_{\adj}\, ,
\hspace{1cm}
\hat{c}_{2}=\delta_{a_{1}a_{3}}\delta_{a_{2}a_{4}} \equiv c_\s \, ,
\label{basis}
\eeq
 which corresponds to octet and singlet for ${\rm SU}(3)$ exchange in the
 $t$-channel and it is the octet that receives the additional suppression
(\ref{suppress}) relative to the singlet.

 In the forward direction, the $\lambda_{_{I}}$, and hence $G$, evidently
 include the leading logarithms in $s$, in the form $\ln(s/t)$.
 This is to be contrasted to the fixed angle case, where the
 leading (double) logs appear as $\ln(Q^2/\Lambda^2)$, with
 $Q^2\sim s$, $t$.  Since $s$-dependence appears in a
 crucial manner in the color-dependence of $G=UH$, while the
 $z_i$'s in eq.\ (\ref{abehave})
 are independent of color flow, it is natural to
 eliminate $s$-dependence from the external self-energies
 altogether.
 This we may accomplish by the choice of gauge
\beq
n^{\mu}=(P_{1}-P_{3})^{\mu}/ \sqrt{-t}=(v_{1}-v_{3})^{\mu}
\sqrt{s/ -2t}\, ,
\label{gauge}
\eeq
 which gives, (see eq.\ (\ref{nudef})),
\beq
\frac{1}{4}\ln\prod_{i=1}^{4}\nu_{i}=-\ln \left( \frac{s}{-t} \right)-\ln2 \, .
\eeq
 Then the eigenvalues become
\beq
\lambda_{\adj}^{(1)} = N_{c} \ln \left(\frac{s}{-t} \right)
-\frac{1}{N_{c}} 2 \pi i + \lambda_{\s}^{(1)}
\mbox{, \hspace{.5cm}}
\lambda_\s^{(1)} = 2 C_{F} + {\cal O}(\ln^{-1}(s/-t)) \, .
\label{eigenvalues}
\eeq
 In LLA with gauge choice (\ref{gauge}), $\lambda_{\adj}^{(1)} \approx
\Delta \lambda^{(1)}$ and we thus find for hard octet exchange
 in eq.\ (\ref{calH}) the combination
\beq
{\cal H}_{\adj}= \left( \frac{s}{-t} \right)^{-\frac{2 N_{c}}{\beta_{1}}
      \ln \left( \frac{\ln(\sqrt{-t}/ \Lambda)}{\ln(1/ b\Lambda)} \right)}
   H_{\adj}(\alpha_s(-t),\theta; \lambda_i) \, ,
\label{regge}
\eeq
 where we again suppress non-leading terms in the exponent,
 and where, at lowest order,
\beq
H_\adj^{(1)} = A_{{\rm Born}} =
\frac{4\pi\alpha_s}{t} (\bar{u}_3 \gamma^\mu u_1)
(\bar{u}_4 \gamma_\mu u_2) \, .
\eeq

 Eq.\ (\ref{regge}) gives the reggeized
 form of the amplitude for hard octet exchange as derived from the fixed angle
 formalism of \cite{BS}. This is to be compared with the standard
 result \cite{Reg1,Reg2} for the amplitude
 from leading logarithms in $s$ only,
\beq
A_{\adj}=\left( \frac{s}{-t} \right)^{\alpha_{g}(-t)}
A_{{\rm Born}} \, ,
\eeq
 where, using notation ${\bf q}$ for $q^\mu_\perp$,
\beq
j_{g}(t) = 1 +\alpha_{g}({\bf q}^{2})=1-\frac{N_{c}\alpha_s}
{4 \pi^{2}} {\bf q}^{2} \int \frac {\d ^{2}{\bf k}} {{\bf k}^{2}
({\bf q}-{\bf k})^{2}},
\hspace{.5cm} {\bf q}^{2}=-t \, ,
\label{alpha}
\eeq
 is the gluon Regge trajectory. The quantity $\alpha_{g}(-t)$ as
 defined in (\ref{alpha}) is IR divergent at ${\bf k} \rightarrow {\bf 0}$ and
 ${\bf k} \rightarrow {\bf q}$. Dimensional regularization
 gives, for $ \alpha_{g}(-t)$,
\beq
\alpha_g ={N_c\alpha_s \over 2\pi}{1\over \e} + {\cal O}(\e^0)\, .
\label{trajdimreg}
\eeq
 The relationship of this leading $\ln (s)$ result with eq.\ (\ref{regge})
 becomes clear when we substitute the octet eigenvalue in
 (\ref{eigenvalues}) into the exponent eq.\ (\ref{gammadef}),
 taking $b \rightarrow \infty$ and
 using the zeroth order (`fixed') running coupling in $D$ dimensions,
\beq
\alpha_s(\mu')=\alpha_s(\mu)(\mu/\mu')^{2\e}\, .
\label{zerocoup}
\eeq
 The result is precisely eq.\ (\ref{trajdimreg}).

 We should note that the correspondence of our eigenvectors
 to singlet and octet exchange is only exact for $s\rightarrow \infty$.
 According to eq.\ (\ref{edef}),
 the singlet-octet basis diagonalizes
 ${\bf \Gamma}_{U}$ only to ${\cal O}(\ln(s/t))$. Indeed
\beq
{\bf R}^{-1} {\bf \Gamma}_{U} {\bf R} = \frac{\alpha_s}{\pi} \left(
\begin{array}{cc}
\lambda_{\adj}^{(1)}             &  2 \pi i \\
\frac{C_{_{F}}}{N_{c}}\pi i    &  \lambda_\s^{(1)}
\end{array}
\right) + {\cal O}(\ln^{-1}(s/t))\, .
\label{mixing}
\eeq
 The off-diagonal elements describe an ${\cal O}(\ln^{-1}(s/t))$ mixing
 of the singlet and the octet in the basis that would diagonalize
 ${\bf \Gamma}_{U}$ to ${\cal O}(\ln^{0}(s/t))$. But in view of the
 octet reggeization, the contribution to the evolution from this mixing
 dies out at high energy and the amplitude is singlet dominated.

 For the singlet vertex function we obtain in place of eq.\ (\ref{suppress})
 a suppression in $t$ only in LLA,
\beq
\exp\left[
           - \frac{2\lambda_\s^{(1)}}{\beta_1}\ln \left(
          \frac{\ln (\sqrt{-t}/ \Lambda)}{\ln (1/ b\Lambda)} \right)
    \right]
  = \left( \frac{\ln (\sqrt{-t}/ \Lambda)}{\ln (1/ b\Lambda)} \right)
    ^{-\frac{4 C_{F}}{\beta_{1}}} \, .
\eeq
 Evidently, for sufficiently large $s$ and fixed $t$, the singlet
 dominates the amplitude and we find from eq.\ (\ref{abehave}) in LLA
 in $s$ and $t$ separately,
\beqa
A_I &= & \exp\left[ - \frac{8 C_{F}}{\beta_1}
\ln \left( \frac{\sqrt{-t}}{\Lambda} \right)
 \ln \left( \frac{\ln (\sqrt{-t}/ \Lambda)}{\ln (1/ b\Lambda)} \right)
        + {\rm NL}     \right]     \nonumber \\
&\ & \times S_{I,\s}(\alpha_s(1/b),\theta,\epsilon)
H_\s(\alpha_s(-t),\theta;\lambda_i) \, ,
\label{evolvesinglet}
\eeqa
 where $H_\s$ is the `hard singlet amplitude' , IR finite and
 independent of all scales below the hard scale $-t$,
 and $S_{I,\s} \equiv S_{I2}$ in the singlet-octet basis
 of eq.\ (\ref{basis}).
 Unlike the case of octet exchange, however, where the lowest order
 contribution is $A_{{\rm Born}}$, for the singlet the lowest possible
 order contribution is ${\cal O}(\alpha_s^{2})$.
 In order to use eq.\ (\ref{evolvesinglet})
 in its perturbative expansion
 we need to construct the  lowest order IR finite  amplitude $H_\s^{(1)}$.

\section{Lowest order singlet exchange}

 In this section we review singlet exchange at one loop.
 For massless q-q elastic scattering in the region
\beq
s \gg -t \gg \Lambda^{2} \, ,
\label{region}
\eeq
 we have seen that it is natural to write the amplitude $A_{ \{ a_{i} \} }$
 in the singlet-octet basis of eq.\ (\ref{basis}),
\beq
A_{ \{ a_{i} \} }(s,t;\lambda_{i}) = A_\s(s,t;\lambda_{i})(c_\s)_{ \{ a_i \}}
+ A_{\adj}(s,t;\lambda_{i})(c_{\adj})_{ \{ a_i \}} \, .
\eeq
 The lowest order contribution, ${\cal O}(\alpha_s^2)$, to $A_\s$
comes only from graphs (a), (b) of fig.\ 3,  whose color coefficients are
\beq
C^{(a)}_\s=C^{(b)}_\s=(N^{2}_{c}-1)/4N^{2}_{c} \equiv C_\s \, ,
\label{colorfactor}
\eeq
\beq
C^{(a)}_{\adj}=-1/N_{c} \mbox{, \hspace{1 cm}}
C^{(b)}_{\adj}=(N^{2}_{c}-2)/2N_{c} \, .
\eeq
 To this order $A_\s^{(1)}$ is gauge invariant, since
 the one-loop vertex and gluon self-energy renormalization
 do not contribute to $A^{(1)}_\s$.
 We perform the calculation in Feynman gauge.
 The graphs (a), (b) are UV finite. Their
 analytic structure is exhibited through the use of the parameter
\beq
z=\frac{-t}{s}-i{\cal \varepsilon} \, .
\eeq
 The one-loop calculation gives for the singlet
 contribution of graphs (a), (b),
\beqa
A^{(a)}_\s &=& C_\s(\alpha_s\mu^{\epsilon})^{2}
      \left( \frac{4 \pi \mu^{2}}{-t} \right) ^{\epsilon}
      \Gamma (2+\epsilon) (-z)^{\epsilon}
      \left\{ \frac{1}{\epsilon^{2}} - \frac{1}{\epsilon} - \frac{1}{2}
      \ln ^{2}(-z) - 4\zeta(2) +1
      \right\}                         \nonumber \\
     &\ & \quad \times  \frac{4}{t}
      ( \bar{u}_{3} \gamma^{\mu} u_{1} )( \bar{u}_{4} \gamma_{\mu} u_{2} )
      +R^{(a)} \, ,
\label{Aa}
\eeqa
\beqa
A^{(b)}_\s &=&- C_\s(\alpha_s\mu^{\epsilon})^{2}
      \left( \frac{4 \pi \mu^{2}}{-t} \right) ^{\epsilon}
      \Gamma (2+\epsilon)z^{\epsilon}
      \left\{ \frac{1}{\epsilon^{2}} -\frac{1}{\epsilon}- \frac{1}{2}
      \ln^{2}z - 4\zeta(2) +1
      \right\}                         \nonumber \\
      &\ & \quad \times \frac{4}{t}
      ( \bar{u}_{3} \gamma^{\mu} u_{1} )
      ( \bar{u}_{4} \gamma_{\mu} u_{2} )
      +R^{(b)} \, .
\label{Ab}
\eeqa
 The remainders $R^{(a)}$ and $R^{(b)}$ are IR finite and power suppressed by
 ${\cal O}(|z|)$.
 $A_\s^{(a)}$, $A_\s^{(b)}$ manifestly satisfy $s$-channel two-particle
 unitarity.
 Before attempting further expansion in $\epsilon$, let us identify
 the origin of the various pieces in eqs.\ (\ref{Aa}), (\ref{Ab}).
 The terms inside the braces come from
 integration over the Feynman parameters, and the $\epsilon$-poles are
 the soft and collinear divergences. The terms
 $\Gamma(2+\epsilon)(\pm z)^{\epsilon}$ come from  integration over the loop
 momentum.  Upon expanding in
 $\epsilon$ and adding the contributions we obtain simply
\beq
A^{(1)}_\s(s,t,\epsilon;\lambda_{i}) = C_\s(\alpha_s\mu^{\epsilon})^{2}
\left( \frac{4\pi\mu^{2}}{-t {\rm e}^{\gamma_E}} \right) ^{\epsilon}
\frac{i\pi}{\epsilon} \frac{4}{t}
      ( \bar{u}_{3} \gamma^{\mu} u_{1} )
      ( \bar{u}_{4} \gamma_{\mu} u_{2} ) \, .
\eeq
For the renormalization scale, the $\overline{\rm MS}$ choice
$\mu^{2}= \overline{Q}^{2}$, with
\beq
\overline{Q}^{2} \equiv \frac{\exp (\gamma_{_{E}})}{4\pi}(-t) \, ,
\eeq
 sets the scale of the hard scattering and is compatible with the
 one-loop gluon self-energy and vertex renormalization that contribute to
 $A_{\adj}$ to ${\cal O}(\alpha_s^{2})$ ~\cite{Ellis}.
 Then the amplitude becomes
\beq
A^{(1)}_\s(s,t,\epsilon;\lambda_{i}) = C_\s
             (\alpha_s\mu^{\epsilon})^{2}
             \frac{i\pi}{\epsilon} \frac{4}{t}
             ( \bar{u}_{3} \gamma^{\mu} u_{1} )
             ( \bar{u}_{4} \gamma_{\mu} u_{2} )
  = C_\s \frac{i\alpha_s}{\epsilon} A_{{\rm Born}} \, .
\label{As1MS}
\eeq
 The $\overline{\rm MS}$ choice for the hard scale can be easily included
 in eq.\ (\ref{evolvesinglet}) by reinstating the $C_2$ parameter as in
 eq.\ (\ref{evolsoln}) and setting
 $(C_{2})^{2}=\exp(\gamma_{_{E}})/ 4 \pi$.
 For any other scale choice $\mu^{2}$ the amplitude becomes
\beq
A^{(1)}_\s(s,t,\epsilon;\lambda_{i}) = C_\s
            i\alpha_s
           \left(  \frac{1}{\epsilon}+\ln\frac{\mu^{2}}{\overline{Q}^{2}}
           \right) A_{{\rm Born}} \, .
\label{As1}
\eeq

 The leading IR singularities $1/ \epsilon^{2}$ cancel in the singlet
 exchange. We observe that $A^{(1)}_\s$ in region (\ref{region}) is
 purely imaginary to leading power in $s/t$ and that there are
 no IR regular terms in $\overline{\rm MS}\;$scheme.
 All these features  are well known. We will now discuss the
 IR subtraction procedure for the above amplitude.

\section{IR subtractions}

 The IR singular part of $A^{(1)}_\s$ comes from soft gluon exchange.
 The soft region in loop momentum space
 is defined in light cone coordinates as the region where gluon momenta
 scale as
\beq
q_{1}^{\mu}=(q_{1}^{+},q_{1}^{-};q_{1 \perp}) \approx
Q (\lambda,\lambda;\lambda) \, ,
\label{softscaling}
\eeq
 with $\lambda \rightarrow 0$. Although collinear singularities are present
 in (a), (b) in the Feynman gauge, unlike in axial gauge, their net effect
 cancels in singlet exchange.

 The contributions to the one-loop singlet amplitude, factorized as
 in eq.\ (\ref{factorizationG}), are
\beq
A^{(1)}_\s = U^{(1)}_{\s,\adj}(\alpha_s(\mu),\theta,\epsilon) A_{\rm Born}
 + H_\s^{(1)} (Q/ \mu, \alpha_s(\mu),\theta;\lambda_i) \, .
\label{splitA}
\eeq
 As noted above $U_{\s,\adj}$ is the singlet-octet mixing term, whose
 {\em unrnormalized} value is zero in dimensional regularization
 (see eq.\ (\ref{uequalz})). We may then write
\beqa
U_{\s,\adj}^{(1)} &=&  C_\s
 (\omega^{(a)} + \omega^{(b)})_{\rm ren} \, , \nonumber \\
H_\s^{(1)} &=& A_\s^{(1)} -  C_\s
 (\omega^{(a)} + \omega^{(b)})_{\rm ren} A_{ {\rm Born}} \, .
\label{UHsub}
\eeqa
 Here the $\omega^{(a,b)}$ are given by the eikonal graphs of fig.\ 4,
\beqa
\omega^{(a)} &=& 4\pi\alpha_s\mu^{2\epsilon}2 \int
\frac{\d ^{D}q_{1}}{(2\pi)^{D}} \frac{(-i)v_{1}\cdot v_{2}}
{(q_{1}^{2}+i\epsilon) (-v_{1} \cdot q_{1} +i \epsilon)
(v_{2} \cdot q_{1} +i \epsilon)} \, , \nonumber \\
\omega^{(b)} &=& 4\pi\alpha_s\mu^{2\epsilon}2\int
\frac{\d ^{D}q_{1}}{(2\pi)^{D}} \frac{(-i)v_{1}\cdot v_{4}}
{(q_{1}^{2}+i\epsilon) (-v_{1} \cdot q_{1} +i \epsilon)
(-v_{4} \cdot q_{1} +i \epsilon)} \, .
\label{omegas}
\eeqa
 The factor of 2 is due to equal contributions from left-right mirror
 graphs. The subscript `ren' in eq.\ (\ref{UHsub}) indicates that
 $\omega^{(a,b)}$ vanish in $D$ dimensions and are {\em defined } by a
 renormalization prescription that uncovers their IR poles at $D=4$.
 Thus we must evaluate the integrals formally to identify an IR pole term,
 which will serve as the one-loop UV counterterm, $(Z_{_U}^{-1})^{(1)}$,
 as in eq.\ (\ref{uequalz}). To facilitate this process,
 we multiply and divide the integrands in
 $\omega^{(a,b)}$ by $\mu^2 / \kappa^2$, with $\kappa$
 a dimensionless free parameter and use Feynman parametrization on the
 denominators  $\pm (\mu/ \kappa)  v_i \cdot q_1 + i \epsilon$.
 This procedure gives the
 formal expressions for the unrenormalized integrals:
\beqa
\omega^{(a)} &=&  \frac{\alpha_s}{\pi}
    \left( \frac{8 \pi \kappa^2}{-v_1 \cdot v_2} \right)^\epsilon
    \Gamma(1+ \epsilon)
    B(2\epsilon, -2\epsilon) B(-\epsilon,-\epsilon) \, , \nonumber \\
\omega^{(b)} &=& -\frac{\alpha_s}{\pi}
     \left( \frac{8 \pi \kappa^2}{ v_1 \cdot v_4} \right)^\epsilon
    \Gamma(1+\epsilon)
    B(2\epsilon, -2\epsilon) B(-\epsilon,-\epsilon) \, .
\label{calcomega}
\eeqa
 Of course, since $B(2\epsilon, -2\epsilon) = 0$, these forms are not unique.
 The UV pole, however, may be isolated by applying the identity
\beq
B(a,b) =   B(1+a,b)+B(a,1+b)
\label{Bid}
\eeq
 to $B(2\epsilon,-2\epsilon)$ and identifying poles from $\Gamma(n\epsilon)$
 as UV (convergent for $\epsilon > 0$ ) and  $\Gamma(-n\epsilon)$ as IR
 (convergent for $\epsilon < 0$ ). In this fashion, we identify purely
 IR pole terms for the $\omega$'s, which we denote by
 $\omega^{(a,b)}_{\rm ren}$:
\beqa
\omega^{(a)}_{\rm ren} &=&  \frac{\alpha_s}{\pi}
         \left( \frac{8 \pi \kappa^2}{-v_1 \cdot v_2} \right)^\epsilon
         \Gamma(1+\epsilon)
  \left\{
  \frac{1}{\epsilon^2} + 3 \zeta (2)+ {\cal O}(\epsilon)
  \right\} \, , \nonumber \\
\omega^{(b)}_{\rm ren} &=& - \frac{\alpha_s}{\pi}
        \left( \frac{8 \pi \kappa^2}{v_1 \cdot v_4} \right)^\epsilon
        \Gamma(1+\epsilon)
  \left\{
  \frac{1}{\epsilon^2} + 3 \zeta (2)+ {\cal O}(\epsilon)
  \right\} \, .
\label{renomega}
\eeqa
 Here our choice of $\kappa^2$ will determine the precise renormalization
 of $U_{\s,\adj}$. $U_{\s,\adj}$ must be chosen to absorb the IR divergences
 of $A_\s^{(1)}$, eqs.\ (\ref{As1}), (\ref{As1MS}),
 but is otherwise arbitrary. Its general form at one loop is
\beqa
U_{\s,\adj}^{(1)} &=& C_\s [ \omega^{(a)} + \omega^{(b)} ]_{\rm ren}
                  \nonumber \\
                &=& C_\s \frac{\alpha_s}{\pi} \frac{1}{\epsilon^2}
               [ (-v_1 \cdot v_2)^{-\epsilon} - (v_1 \cdot v_4)^{-\epsilon} ]
               [ 1 + \epsilon ( \ln (8 \pi \kappa^2) - \gamma_{_E} )]
               + {\cal O}(\epsilon)  \nonumber \\
         &=& C_\s \frac{\alpha_s}{\pi}
            \left\{
   \frac{i \pi}{\epsilon} +
   \frac{ \ln ( v_1 \cdot v_2 / v_1 \cdot v_4) }{\epsilon} +
   i \pi [ \ln (8 \pi \kappa^2)- \gamma_{_E} ] - \frac{\pi^2}{2}
           \right\}+ {\cal O}(\epsilon)\, .\nonumber
	   \\
\label{UIR}
\eeqa
 We may choose $\kappa^2$ to eliminate all or part of the finite terms that
 accompany the imaginary IR divergence in $A_\s^{(1)}$.
 (Note that $v_1 \cdot v_2 / v_1 \cdot v_4 =1$ in the approximation
 $ -t/s \ll 1$, in which we work.) In a strictly `minimal' scheme, we might
 well choose $\kappa^2$ to cancel all the finite terms in $U_{\s,\adj}^{(1)}$,
 so that $U_{\s,\adj}^{(1)} A_{\rm Born} = A_\s^{(1)}$. In this case,
 we would have, by eq.\ (\ref{splitA}), $H_\s^{(1)}=0$ for $\overline{MS}$
 scheme $\mu = \overline{Q}$. This choice is not
 unique, however. The optimal choice should be guided  by the manner
 in which the factorized amplitude is embedded in an infrared-safe
 quantity. In such a quantity  we may expect the IR divergences of $U$
 to cancel. If other, finite, contributions to $A_\s$ cancel as well,
 these may naturally be included in $U$, at least if they arise from regions
 of `soft' loop momentum, as in eq.\ (\ref{softscaling}),
 with $0 < \lambda < 1$.

 From the above considerations, we may identify an alternate definition of
 $\kappa^2$ in eq.\ (\ref{UIR}),
\beq
\kappa^2 = \frac{1}{8 \pi} {\rm e} ^{\gamma_{_E}} \, ,
\label{kappa}
\eeq
for which
\beqa
U_{\s,\adj}^{(1)} &\equiv&  C_\s \frac{\alpha_s}{\pi}
          \frac{1}{\epsilon^2}
           [ (-v_1 \cdot v_2)^{-\epsilon} - (v_1 \cdot v_4)^{-\epsilon} ]
          \nonumber \\
       &=&  C_\s \frac{\alpha_s}{\pi}
          \frac{1}{\epsilon^2} [(-1)^\epsilon -1] \, .
\label{Ufix}
\eeqa
 That is, we keep ${\cal O}(\epsilon^0)$ terms that follow directly from the
 combination of the double IR pole with the analytic behavior of the graphs as
 functions of their external momenta. We conjecture that in the calculation
 of the proton-proton elastic scattering amplitude, the entire $U_{\s,\adj}$
 contribution of this form {\em cancels} in the sum over soft corrections.
 When the IR function $U_{\s,\adj}^{(1)}$ is defined according to the latter
 procedure, we find the result
\beq
U_{\s,\adj}^{(1)} A_{\rm Born}  = C_\s (\alpha_s \mu^{\epsilon})^{2}
             \frac{4}{t}
             \left( \frac{i\pi}{\epsilon}-\frac{\pi^{2}}{2} \right)
             ( \bar{u}_{3} \gamma^{\mu} u_{1} )
             ( \bar{u}_{4} \gamma_{\mu} u_{2} ) \, .
\label{UAwithpi}
\eeq
 So, by eqs.\ (\ref{UHsub}) and (\ref{As1}), the lowest order
 IR finite amplitude for  singlet exchange at scale $\mu^2$ is
\beqa
H_\s^{(1)}(Q/ \mu, s/t;\lambda_i) &=& C_\s (\alpha_s\mu^{\epsilon})^{2}
         \left( i \pi \ln \frac{\mu^{2}}{\overline{Q}^{2}}+
\frac{\pi^{2}}{2} \right) \frac{4}{t}
 ( \bar{u}_{3} \gamma^{\mu} u_{1} )( \bar{u}_{4} \gamma_{\mu} u_{2} )
\nonumber \\
 &=& C_\s \alpha_s \left( i\ln \frac{\mu^{2}}{\overline{Q}^{2}}+
\frac{\pi}{2} \right) A_{{\rm Born}} \, .
\label{seed}
\eeqa
 Notice again that had we not kept the ${\cal O}(\epsilon^0)$ terms in
 eq.\ (\ref{Ufix}), the $\overline{\rm MS}$ choice $\mu=\overline{Q}$
 would have given $H_\s^{(1)}=0$. The real part of $H_\s^{(1)}$ is
 generated by the above oversubtraction procedure.

 \section{Summary}

 In this paper, we have found that
 in near-forward q-q elastic amplitude with large momentum
 transfer, the hard singlet component
 evolves at leading logarithm in $t$
 according to  eq.\ (\ref{evolvesinglet}).
 The singlet
 hard-scattering function $H_\s(s/t;\lambda_{i})$
 is by construction free from all scales
 below the hard scale $t$.  We have observed that the
 value of the hard-scattering function depends on the
 infrared safe quantity in which the amplitude appears.
 One possibility for the
 $H_\s(s/t;\lambda_{i})$ at one loop is given in  eq.\ (\ref{seed}).
 The singlet evolution shown
 in eq.\ (\ref{evolvesinglet}) organizes all leading and non-leading Sudakov
 $\ln(t)$ behavior.
 The basis of this derivation is the factorization of the
 amplitude as shown in eqs.\ (\ref{qqamp}) and (\ref{factorizationG}).

 In future work we shall examine the $s$-dependence of
 $H_\s$ beyond the
 lowest order in $\alpha_s$. This dependence
 cannot be captured by the Sudakov-oriented approach
 of ~\cite{BS}.
 We anticipate that the leading logarithmic
 $s/t$ dependence of $H_\s$
 is given by a `BFKL' equation \cite{Reg2}.
 Of interest is the possible influence of infrared subtractions on
 the resulting evolution \cite{Mu}.
 In addition, we hope to pursue applications to
 both proton-proton elastic scattering and to jet production.
 For proton-proton elastic scattering, our results
 indicate an important role for a perturbative singlet
 exchange in the independent scattering of quarks \cite{Land}.

\vspace{.5cm}
 {\em Acknowledgements}. We wish to thank I.\ Balitsky,
 V.\ del Duca, I.\ Korchemskaya, G.\ Korchemsky, A. H. Mueller and
 W.-K. Tang for helpful conversations.
 This work is supported in part by the National Science Foundation under
 grant  PHY 9309888 and by the Texas National Research Laboratory.

\newpage

\begin{center}
{\bf Figure Captions}
\end{center}

\begin{enumerate}
\item  Leading radiative corrections to fixed angle quark-quark
      elastic scattering in the axial gauge.

\item  Factorized form of the 4-point on-shell vertex function $G$.
      The internal lines of $H$ are off-shell by ${\cal O}(Q^2)$.

\item  Lowest order graphs for singlet exchange in the $t$-channel.

\item  Eikonal approximation applied to the graphs of fig.\ 3.

\end{enumerate}

\end{document}